\documentclass{aa}  
\usepackage{graphicx}
\usepackage{txfonts}
\begin{document}
   \title{Upper limits to the water abundance in starburst galaxies\thanks{Based
  on observations with Odin, a Swedish-led satellite project funded jointly by
  the Swedish National Space Board (SNSB), the Canadian Space Agency (CSA),
  the National Technology Agency of Finland (Tekes)  and Centre National
  d'Etude Spatiale (CNES). The Swedish Space Corporation has been the 
industrial
  prime contractor and also is operating the satellite.}}


   \author{C. D. Wilson\inst{1}
\and
          R. S. Booth\inst{2,10}
\and
	  A. O. H. Olofsson\inst{2}
\and
          M. Olberg\inst{2}
\and
	  C. M. Persson\inst{2}
\and
          Aa. Sandqvist\inst{3}
\and
	  \AA. Hjalmarson\inst{2}
\and
	  V. Buat\inst{4}
\and
	  P. J. Encrenaz\inst{5}
\and
	  M. Fich\inst{6}
\and
	  U. Frisk\inst{7}
\and
	  M. Gerin\inst{8}
\and
	  G. Rydback\inst{2}
\and
	  T. Wiklind\inst{2,9}
          }

   \offprints{C. Wilson}

   \institute{Department of Physics \& Astronomy, McMaster University,
                Hamilton, Ontario, L8S 4M1 Canada\\
              \email{wilson@physics.mcmaster.ca}
         \and
             Onsala Space Observatory, SE-439 92, Onsala Sweden
	 \and
	     Stockholm Observatory, AlbaNova University Center, SE-106 91 Stockholm, Sweden
	 \and
	 Laboratoire d'Astronomie Spatiale, BP 8, 13376 Marseille Cedex 12, France
	 \and
LERMA \& UMR 8112 du CNRS, Observatoire de Paris, 61, Av. de l'Observatoire, 75014 Paris, France
	 \and
Department of Physics, University of Waterloo, Waterloo, ON N2L 3G1, Canada
	 \and
        Swedish Space Corporation, P O Box 4207, SE-171 04 Solna, Sweden
	 \and
LERMA \& UMR 8112 du CNRS, \'Ecole Normale Sup\'erieure, 24 rue Lhomond, 75005 Paris, France
	 \and
ESA Space Telescope Division, STScI, 3700 San Martin Drive Baltimore, MD 21218, USA
	 \and
Hartebeesthoek Radio Astromomy Observatory, Box 443, Krugersdorp,
1740, South Africa
             }

   \date{Final version}

 
  \abstract
   {}
   {We have searched for emission from the 557 GHz {\it ortho}-water line in
the interstellar medium of six nearby starburst galaxies.
}
   {We used the Odin satellite to observe the $1_{10}-1_{01}$  transition of
{\it o}-H$_2$O in the galaxies NGC253, IC342, M82, NGC4258, 
CenA, and M51. None of the galaxies in our sample was detected.
} 
   {We derive three sigma 
upper limits to the H$_2$O abundance relative to H$_2$
ranging from $2\times 10^{-9}$ to $1\times 10^{-8}$.
 The best of these upper limits are comparable to the
measured abundance of H$_2$O in the Galactic star forming region W3.
However, if only 10\% of the molecular gas is in very dense
cores, then the water abundance limits in the cores themselves would be larger
by a factor of 10 i.e. $2\times 10^{-8}$ to $1\times 10^{-7}$. }
   {These observations suggest that detections of H$_2$O emission 
in galaxies with the upcoming Herschel Space Observatory 
are likely to require on-source integration times of an hour or more
except in the very brightest extragalactic targets such as M82 and NGC253.
}


   \maketitle
%

\section{Introduction}

Recent space-based observations by SWAS (Melnick et al.
\cite{m00}) and Odin (Frisk et al.
\cite{f03}) have
revealed a low gas-phase abundance of water under most conditions
in the interstellar medium (e.g. Snell et al. \cite{snell00a}, 
Olofsson et al. \cite{o03}). 
This low abundance is most likely
due to the water molecules freezing out to form water ice on dust grains
(e.g. Bergin et al. \cite{b00}). However, gas phase water abundances
100 to 10,000 times larger than the typical ambient cloud value
have been seen toward hot cores with ISO (e.g. van Dishoeck
\& Helmich \cite{vdh96}) and in shock-heated outflows (Bergin et al.
\cite{b03}) and supernova remnants (Snell et al. \cite{s05}). The
significant dependence of the gas-phase water abundance on
environment makes it interesting to search for water emission lines
in even more extreme environments, namely the nuclei of nearby starburst
galaxies.

In this paper, we present sensitive upper limits to the
$1_{10}-1_{01}$ emission line of {\it ortho}-H$_2$O at 556.936 GHz 
in six nearby starburst galaxies. The observations and data
reduction are discussed in Sect.~\ref{obs}. We combine the H$_2$O
upper limits with published CO $J$=1-0 data to derive upper limits
to the H$_2$O abundance relative to H$_2$, $x(o{\rm -H_2 O})$,
in Sect.~\ref{limits}. We summarize and discuss the possibilities
for future H$_2$O detections with the Herschel Space Observatory
in Sect.~\ref{future}.


\section{Observations and Analysis}\label{obs}


   \begin{figure*}
   \centering
\includegraphics[width=6cm,angle=90]{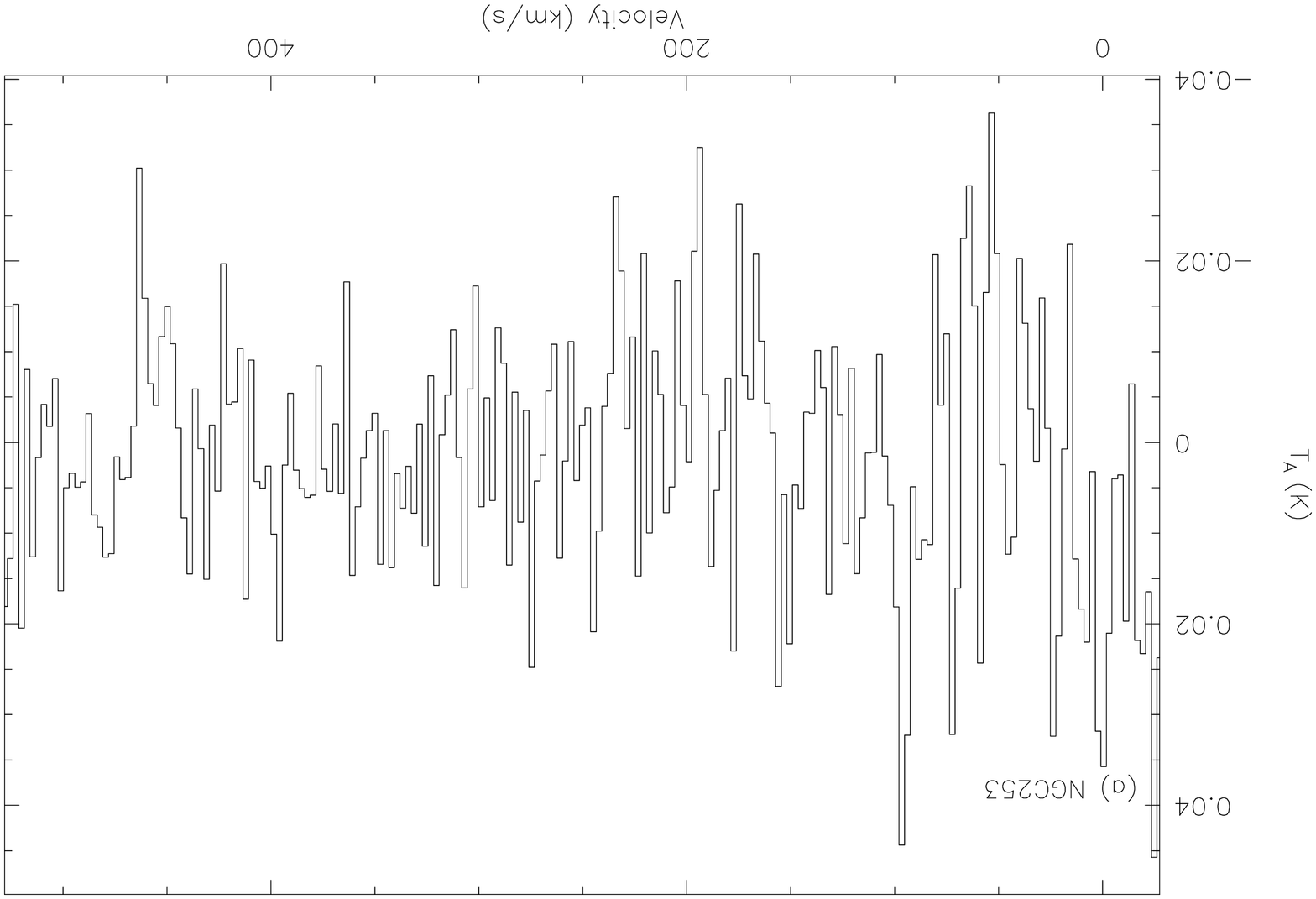}
\includegraphics[width=6cm,angle=90]{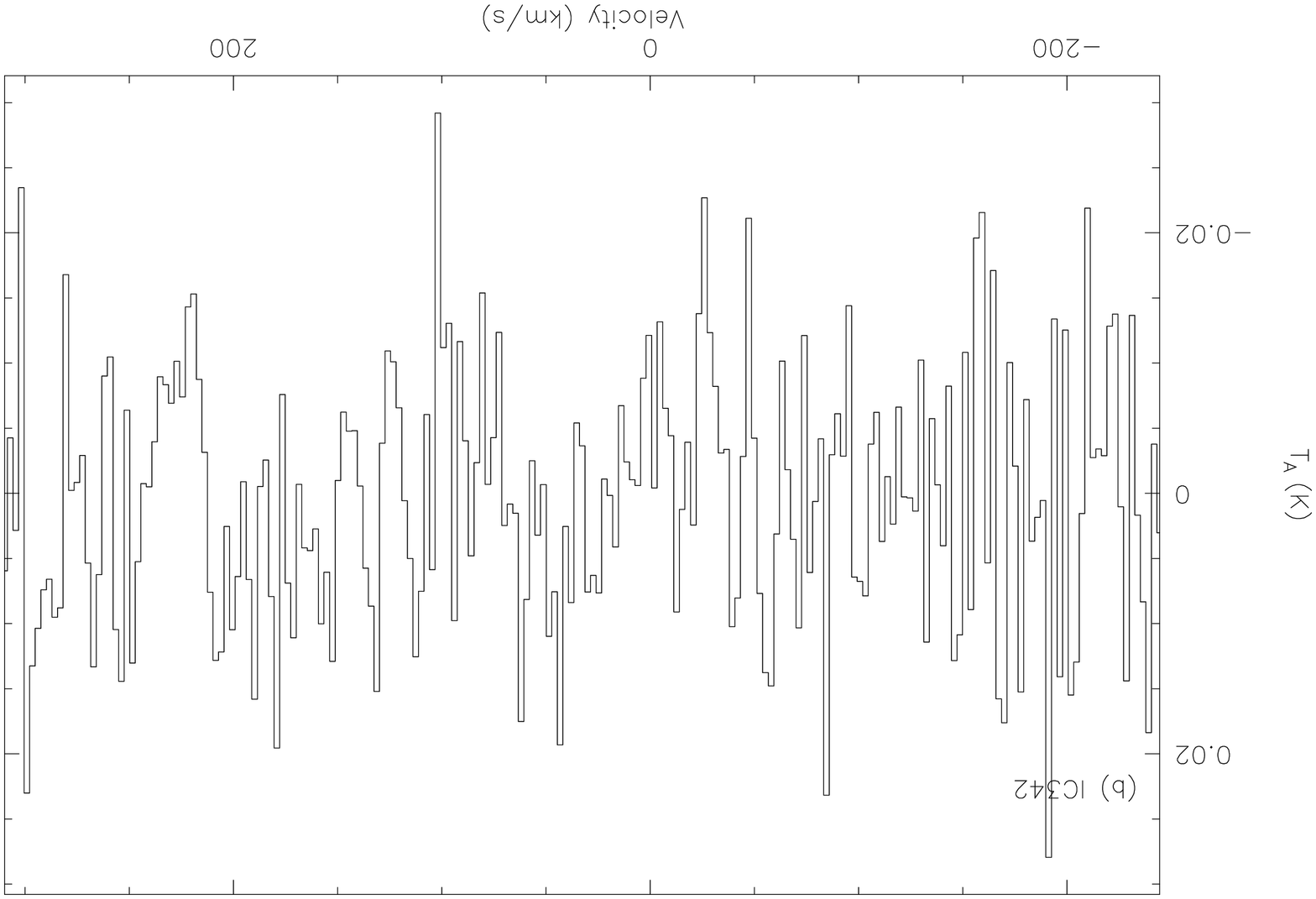}
\includegraphics[width=6cm,angle=90]{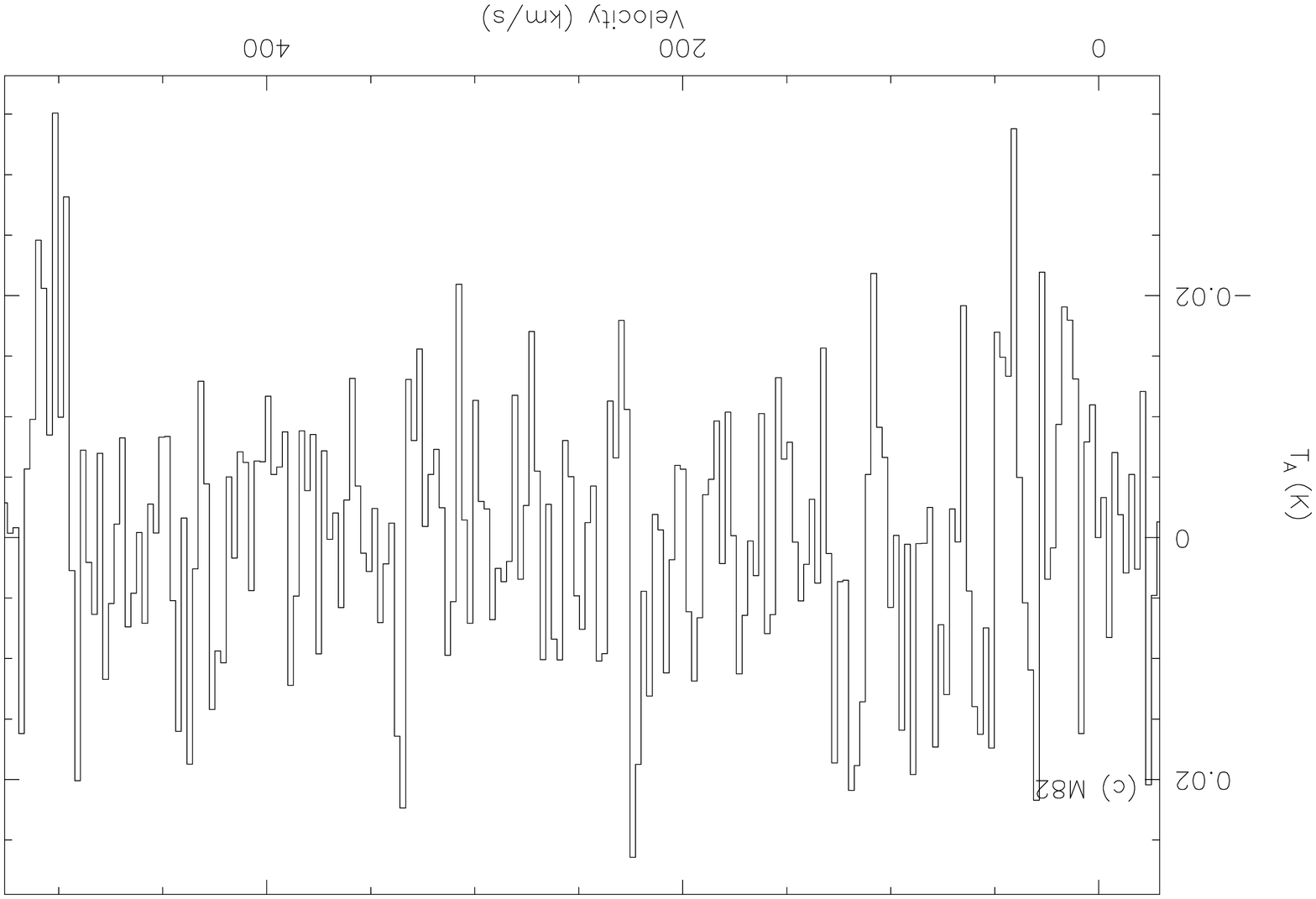}
\includegraphics[width=6cm,angle=90]{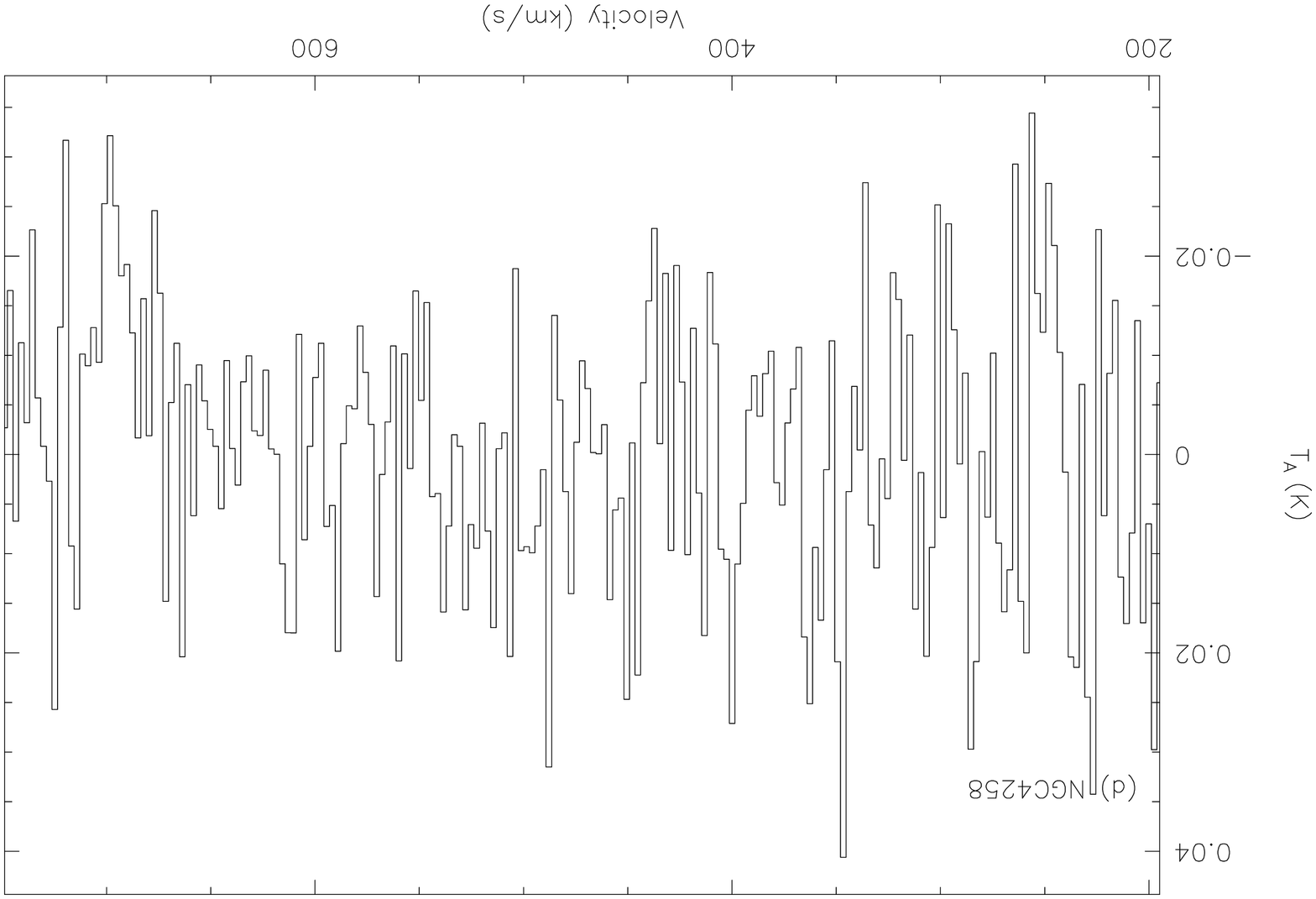}
\includegraphics[width=6cm,angle=90]{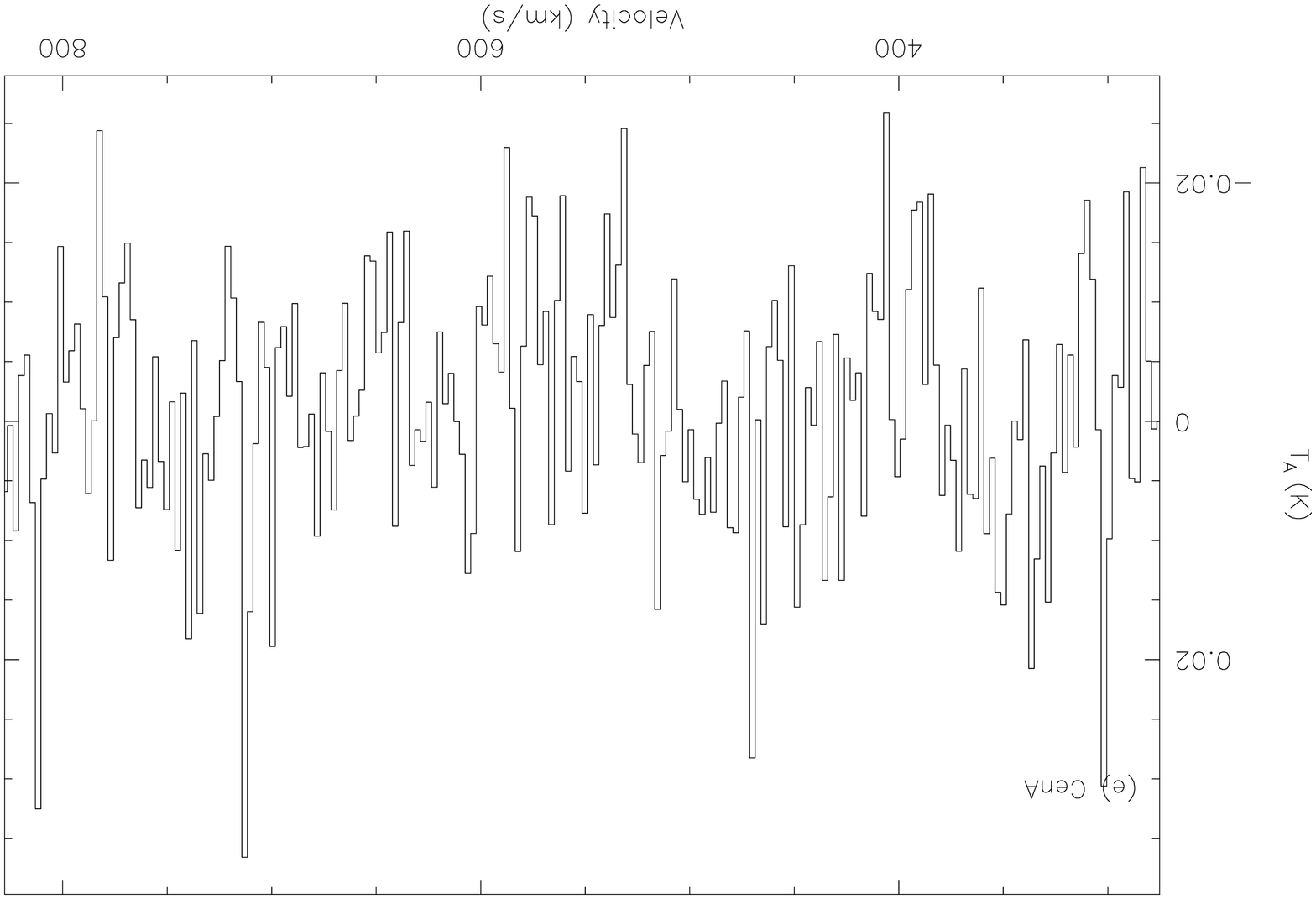}
\includegraphics[width=6cm,angle=90]{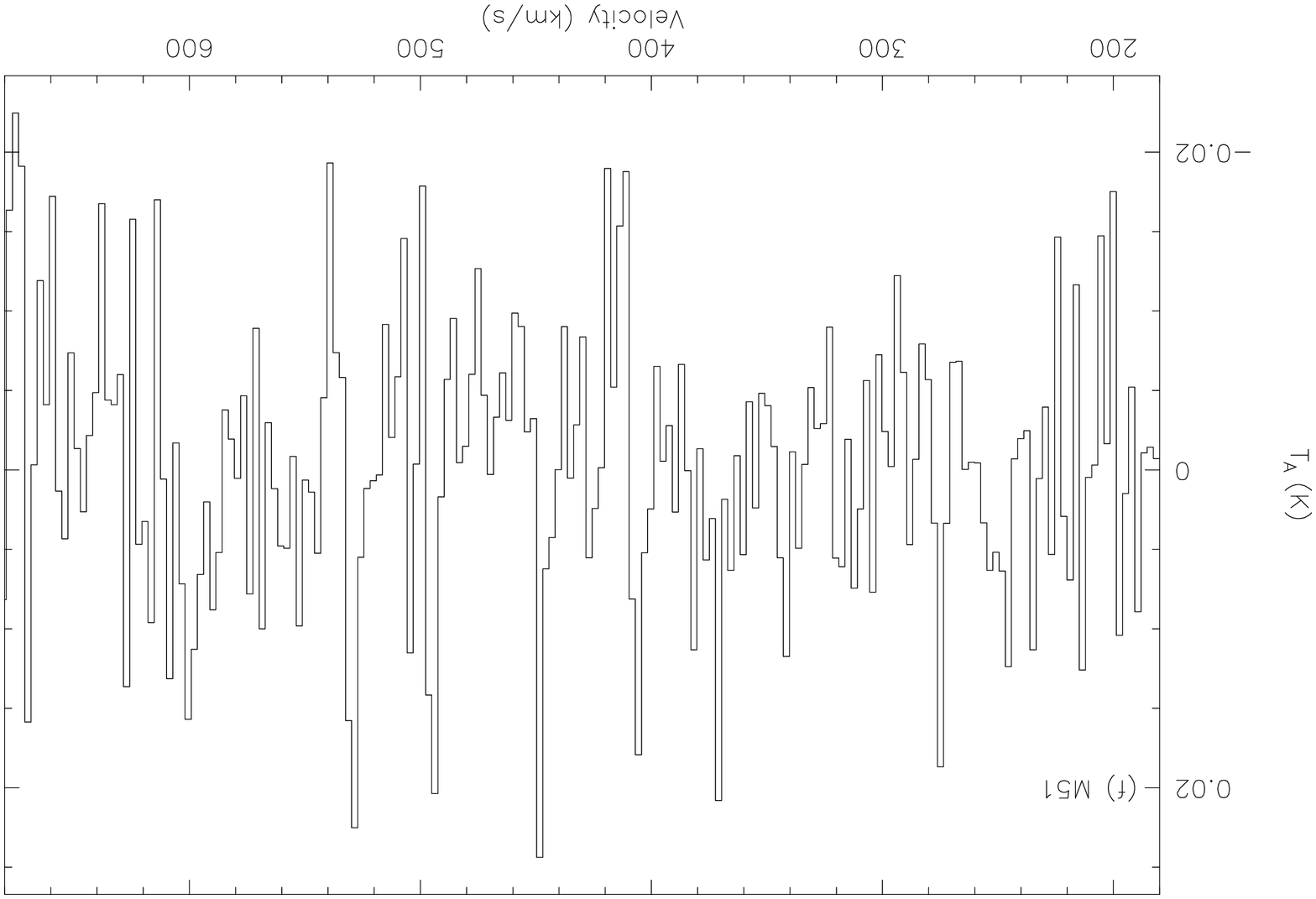}
      \caption{Spectra of nearby starburst galaxies 
tuned to the $1_{10}-1_{01}$  transition of 
{\it ortho}-H$_2$O at 556.936 GHz. The velocity scale is $V_{LSR}$. 
No broad line (consistent with the observed
CO $J$=1-0 emission from these sources) is seen in any of the spectra.
(a) NGC253 ($V_{LSR}$ = 249 km s$^{-1}$). The spectrum
was obtained toward coordinates 00:47:35.2 -25:17:20 (J2000); 
this position is 28$^{\prime\prime}$
from the starburst nucleus. 
(b) IC342 ($V_{LSR}$ = 32 km s$^{-1}$). 
(c) M82 ($V_{LSR}$ = 250 km s$^{-1}$). Due to drifts in Odin's pointing
model with time, this spectrum was obtained at an offset
of $\sim 55^{\prime\prime}$ from the center of M82, which is 
at 09:55:54.0  +69:40:57 (J2000). 
(d) NGC4258 ($V_{LSR}$ = 472 km
s$^{-1}$). 
(e) CenA ($V_{LSR}$ = 550 km s$^{-1}$).
(f) M51 ($V_{LSR}$ = 460 km s$^{-1}$).
              }
         \label{H2Ofig}
   \end{figure*}


\begin{table*}
\begin{minipage}[t]{18cm}
      \caption[]{H$_2$O Observations and Abundance Limits}
         \label{tbl-table}
\centering
\renewcommand{\footnoterule}{} 
\begin{tabular}{lcccccccccl}
            \hline\hline
            Source      & Coordinates & $T_{sys}$ & $t_{int}$
& rms\footnote{Rms noise at a resolution of 2.7 km s$^{-1}$ } & $\Delta V({\rm CO})$  & $I({\rm H_2O})$\footnote{Three sigma upper limit integrated over $\Delta V({\rm CO}$); values for M82 and NGC253 include scaling by 1.61 and 1.15, respectively, to account for pointing errors; see text} & $D$ & $M_{H_2}$\footnote{Assumes $X_{CO} = 2\times 10^{20}$ H$_2$ cm$^{-2}$ (K km s$^{-1}$)$^{-1}$} &
$x(o{\rm -H_2O})$\footnote{Three sigma upper limit assuming $n_{H_2} =
10^6$ cm$^{-3}$ and $T_K = 40$~ K. Depending on the average density
of the molecular gas on large scales and the mass fraction in dense
cores, these abundance limits could be larger by a factor of $\sim 10$
or more (see text).
} 
& CO Reference \\
& (J2000) & (K) & (hr) & (mK) & (km s$^{-1}$) & (K km s$^{-1}$) & (Mpc) & ($10^9$ M$_\odot$) & & \\
            \hline
NGC253 & 00:47:35.2 $-$25:17:20 &  4100 &   18.5     & 13  & 450  & 1.6
& 2.5 & 0.60 & $<2.0\times 10^{-9}$ & Sorai et al. \cite{s00} \\

IC342  & 03:46:49.6  +68:05:45   &  3700  &  25.0 &  10   & 130 & 0.56 & 3.9 & 0.39 & $<2.6\times 10^{-9}$ & Helfer et al. \cite{h03} \\

M82   & 09:55:54.0  +69:40:57   &  3400  & 20.1  &  10   & 400  & 1.9
& 3.9 & 1.56 & $<1.7\times 10^{-9}$ & Walter et al. \cite{w02} \\

N4258  & 12:18:57.5  +47:18:14    &  4300  &  17.4 & 13   & 450  & 1.4 & 8.1 & 0.81 & $<1.3\times 10^{-8}$ & Helfer et al. \cite{h03} \\

CenA  &  13:25:27.6 $-$43:01:08 &  4500  & 17.2 &  10        & 600  & 1.2 & 4.0 & 0.29 & $<7.8\times 10^{-9}$ & Eckart et al. \cite{e90} \\

M51  &   13:29:53.1 +47:11:48 
   &  3900   & 23.7   & ~~8 &   200      &  0.56  & 7.7 & 1.64 & $<2.4\times 10^{-9}$ & Helfer et al. \cite{h03} \\\\ 
\hline
\end{tabular}
\end{minipage}
\end{table*}


We observed the central regions of six nearby starburst galaxies
(\object{NGC253}, \object{IC342}, \object{M82}, \object{NGC4258},
\object{CenA}, and \object{M51})
with Odin between 2002 November 22 and 2005 June 17 with on-source integration 
times ranging from 17 to 25 hours.
Odin's 1.1 m telescope has a beam diameter
of 2.1$^\prime$ at the 556.936 GHz frequency
of the $1_{10}-1_{01}$ transition of {\it ortho}-H$_2$O
and a main beam efficiency
of 0.9 (Frisk et al. \cite{f03}). 
Details of the observations for each galaxy are given in
Table~\ref{tbl-table}. 
The observations were made
in position switching mode; the off positions were located 850$^{\prime\prime}$
to the south-east for NGC253, IC342, M82, and NGC4258, 
280$^{\prime\prime}$ to the south-east for CenA, and 900$^{\prime\prime}$
to the east for M51.
The backend for all observations was an acousto-optic spectrometer
with a spectral resolution of 1 MHz (0.5 km s$^{-1}$) and
a total bandwidth of 1.1 GHz ($\sim 600$ km s$^{-1}$).
Odin's pointing is estimated to be accurate to
15$^{\prime\prime}$. 
However, for M82, maps of Jupiter made after the data were obtained
suggest that
the pointing for this galaxy was in error by $\sim 55^{\prime\prime}$.
We take this pointing error into account in our analysis below.

The data reduction followed the methods described in
Wilson et al. (\cite{w03}) to identify and remove individual
spectra with anomalous system temperatures or 
contamination by the Earth's atmosphere. The individual spectra
were averaged and the resulting spectrum was 
binned to a spectral resolution 
of 2.7 km s$^{-1}$ for the final analysis.
The summed spectra were fit with a 0-2 order polynomial baseline
excluding the expected velocity range of the H$_2$O emission;
typically a velocity range of 200-400 km s$^{-1}$ was excluded
from the baseline fit. The resulting spectra are shown in
Figure ~\ref{H2Ofig}.

Upper limits (3$\sigma$) to the integrated intensity
of the H$_2$O line were calculated using the width of the CO
$J$=1-0 line observed over the same area as the Odin beam. The 
upper limits to the H$_2$O
integrated intensity and the 
CO line widths are given in Table~\ref{tbl-table}.
CO line widths (full width at zero intensity) 
appropriate to the Odin beam were calculated for three galaxies from
the data in the BIMA SONG survey (Helfer et al. \cite{h03}). For CenA
and M82,
the emission in the maps of Eckart et al. (\cite{e90}) 
and Walter et al. (\cite{w02}), respectively, lies almost
entirely within the Odin beam and so we used all the emission to estimate
the line width. For NGC 253, the full width of the CO line is
defined by the spectrum at the starburst nucleus (Sorai et al. \cite{s00}).
For M82, the integrated intensity is corrected for the pointing
error assuming a compact source and an Odin primary beam of 2.1' (full width
at half maximum). For NGC253, the observed coordinates are 28$^{\prime\prime}$
from the peak emission in the galaxy, and so we also correct the
integrated intensity upper limit for this offset.


\section{Upper Limits to the H$_2$O/H$_2$ Abundance Ratio}\label{limits}

We can use the upper limits to the H$_2$O integrated intensity to
calculate upper limits for the H$_2$O abundance within the
Odin beam using the formula given by Snell et al. (\cite{snell00b}). 
We adopt an H$_2$ volume density of $1\times 10^6$ cm$^{-3}$
and a temperature of 40 K to calculate the H$_2$O column
density upper limits. We adopt a relatively high H$_2$ volume density
in these calculations because the H$_2$O emission most likely
originates in the densest parts of the interstellar medium.
 
We use published CO $J$=1-0 data
to calculate the mass of molecular hydrogen within the Odin beam.
For IC342, NGC4258, and
M51, we integrated the CO flux within the Odin beam from
the publicly available BIMA SONG data (Helfer et al. \cite{h03}).
We adopt a CO-to-H$_2$ conversion factor
$X_{CO} = 2\times 10^{20}$ H$_2$ cm$^{-2}$ (K km s$^{-1}$)$^{-1}$
(Strong et al. \cite{s88})
and use the formula in Wilson \& Scoville (\cite{ws90})
to calculate the molecular gas mass. 
For CenA, we take the molecular gas mass from Eckart et al. (\cite{e90})
and rescale it to our adopted conversion factor and distance.
For M82, we take the molecular gas mass from Walter et al. (\cite{w02})
but adjust the mass in the streamers to our adopted conversion factor.
For NGC253, we adopt the mass in the bar plus the ``nuclear bar'' from
Sakai et al. \cite{s00} and scale it to our adopted distance and
conversion factor. The molecular hydrogen gas masses and distances are given
in Table~\ref{tbl-table}, along with the resulting upper limits
for the {\it ortho}-H$_2$O abundance, $x(o{\rm -H_2O})$.

Our limits on the {\it ortho}-H$_2$O abundance 
range from $2\times 10^{-9}$ to $1\times 10^{-8}$. Note that
these estimated abundance limits are inversely proportional to our
assumed value of $10^6$ cm$^{-3}$ for the H$_2$ volume density.
The tightest
constraints are for the four nearest and/or most gas-rich galaxies:
NGC 253, IC342, M82, and M51. For NGC4258, its larger distance is the 
primary reason why the abundance limit is larger than the other
three galaxies, since its molecular hydrogen mass translates into
a smaller average column density in the Odin beam. For CenA, the
smaller molecular hydrogen column density is also the primary
reason for its larger abundance limit.

Goicoechea et  al. (\cite{g05}) have detected H$_2$O in absorption
towards NGC 253 using the Infrared Space Observatory (ISO). 
They derive a lower limit to the H$_2$O column
density of $\ge 1 \times 10^{15}$ cm$^{-2}$. This limit is
significantly larger than the 3$\sigma$ upper limit to the H$_2$O
column density derived from our Odin data, which is $< 4\times
10^{13}$ cm$^{-2}$. This apparent discrepancy can be understood by
examining the regions probed by the two instruments: the Odin upper
limit is an average over a 2.1$^\prime$ region, while the ISO
absorption spectrum probes the much smaller region of the central
continuum source. The central source has a size of $\sim
12^{\prime\prime}$ at mid-infrared wavelengths (F\"orster Schreiber
et al \cite{fs03}) while 
examining archival 850 $\mu$m continuum images from the
James Clerk Maxwell Telescope suggests the size of the central
submillimeter source is no more than 30$^{\prime\prime}$. These
sizes correspond to filling factors of 0.01-0.06 in the larger Odin
beam, which are sufficient to bring the two column density limits
into agreement. Alternatively, assuming a density of
$10^4$ cm $^{-3}$ 
(Bayet et al. \cite{b04}, G\"usten et al. \cite{g06}) would also
bring the two abundance estimates into good agreement (see below).

These water abundance upper limits correspond to values averaged over
kiloparsec scales that, in our own Galaxy, would include many giant
molecular clouds. If we were to picture the molecular interstellar
medium in these starburst galaxies as being similar in structure to
normal spiral galaxies, then only a fraction of the molecular gas
would be contained in cores that are dense enough to produce
significant H$_2$O emission. In this situation, the upper limit to
the water abundance in the cores themselves would be larger than the
values given in Table~\ref{tbl-table}, perhaps by factors of 10-50
depending on the exact fraction of gas in dense cores\footnote{For 
example, the W3 cloud
contains a total of 10$^5$ M$_\odot$ (Lada et al. \cite{l78}), of which 
1900 M$_\odot$ (Tieftrunk et al. \cite{t95}) is contained within the
single core studied in H$_2$O 
emission by Wilson et al. (2003). Depending on the masses of the three
other large cores in W3(OH) and W3 Main (Tieftrunk et al. \cite{t95,t98}), 
this cloud likely contains 2-8\% of its mass with densities high
enough to excite water emission. }. Alternatively,
the molecular interstellar medium of starburst galaxies is likely to have
considerably higher average densities than those of normal galaxies.
If water emission arises from an extended medium with a uniform density of
$10^4-10^5$ cm$^{-3}$ (Bayet et al. \cite{b04}, G\"usten et al. \cite{g06}), 
the upper limits on large scales given in Table~\ref{tbl-table} would
increase by factors of 10-100.

It is interesting to compare these upper limits to the H$_2$O abundance
in starburst galaxies with the H$_2$O abundance estimated for
Galactic star forming regions. 
Snell et al. (\cite{snell00a}) estimate H$_2$O abundances
ranging from $6\times 10^{-10}$ to $1\times 10^{-8}$ for
eight giant molecular cloud cores. 
(They also obtain upper limits
for the H$_2$O abundance in dark clouds, with the most stringent upper
limit of $7\times 10^{-8}$ being obtained for TMC-1.)
Wilson et al. (\cite{w03}) obtain an H$_2$O abundance for the W3 IRS5
region of $2\times 10^{-9}$. 
Olofsson et al. (\cite{o03}) used Odin to detect H$_2^{18}$O at two
positions in Orion and derive an H$_2$O abundance of
$\sim 10^{-5}$ toward the outflow associated with Orion BN-KL and
$(1-8)\times 10^{-8}$ in the ambient cloud 2$^\prime$ to the south.
More recently, Wirstr\"om et al. (\cite{w06}) have used C$^{18}$O $J$=5-4
and H$_2$O observations from Odin to measure a water abundance of
$\ge 8\times 10^{-8}$ in the Orion photon-dominated region.
Plume et al. (\cite{p04}) have used absorption line measurements toward
W49A to obtain H$_2$O abundances of $8\times 10^{-8}$ to $4\times 10^{-7}$
toward three foreground clouds.  Bergin et al. (\cite{b03}) found much higher abundances 
of $> 10^{-6}$ toward three outflows in the NGC 1333 molecular cloud core.
Snell et al. (\cite{s05}) have detected H$_2$O emission in the
supernova remnant IC443 with abundances (assuming [CO]/[H$_2$] = $10^{-4}$)
between  $2\times 10^{-8}$ and $3\times 10^{-7}$. They suggest that
both photodissociation of H$_2$O in post-shock gas and freeze-out of
H$_2$O in the ambient gas are needed to explain the weakness of the
H$_2$O lines.

The above discussion shows that
the Odin upper limits for the H$_2$O abundance averaged over
kiloparsec scales are comparable
to the lowest abundances obtained in the cores of Galactic giant molecular
clouds. Naively, one might expect the interstellar medium in intense
starburst galaxies like M82 to have similar physical conditions to
high-mass star forming cores but with those conditions now present on
kiloparsec scales. The H$_2$O upper limits obtained here suggest that
this naive picture is not correct, and that even in intense starburst galaxies
there is a substantial mass of molecular gas that is not in the warm
dense conditions appropriate to produce H$_2$O emission. 
This conclusion is consistent with recent estimates of the average
gas density of $10^4-10^5$ cm$^{-3}$ in some nearby starburst galaxies
(Bayet et al. \cite{b04}, G\"usten et al. \cite{g06}). 
It is also possible that the intense ultraviolet radiation fields produced in
starburst galaxies photo-dissociate more of the H$_2$O molecules than
is the case for individual high-mass cores like W3. Melnick et al.
(\cite{m05}) suggest that H$_2$O in the gas phase is restricted to a thin
layer of gas in molecular cloud cores, deep enough that it is shielded
from the ambient ultraviolet radiation field but not so deep that
the molecules freeze out onto dust grains. The stronger ultraviolet
radiation fields in starburst galaxies might conceivably 
alter this delicate balance and make H$_2$O an even rarer molecule in the
gas phase than it is in our own Galaxy. Finally, dramatic 
H$_2$O absorption features are seen in observations toward Sgr A
(Sandqvist et al. \cite{s03}) and Sgr B2 (Neufeld et al. \cite{n03})
in the Galactic Center region. These observations 
may imply that self-absorption
of the H$_2$O line could make it very difficult to detect emission
in regions such as starburst nuclei 
with warm dense gas containing multiple outflows
and shocks from supernova remnants.  All these effects may conspire
to make the emission lines of H$_2$O very difficult to detect
even in the brightest and closest starburst galaxies.


\section{Conclusions and Future Prospects}\label{future}

We have obtained long integrations with the Odin satellite in an
attempt to detect the emission line of {\it ortho}-water in
six nearby starburst galaxies.
None of the galaxies were detected, with upper limits to the
H$_2$O integrated intensity ranging from 0.6 to 1.9 K km s$^{-1}$.
We have combined these upper limits with published CO data to
derive upper limits to the H$_2$O abundance in each galaxy. Our
3$\sigma$ upper limits to [H$_2$O]/[H$_2$]
range from $2\times 10^{-9}$ to $1\times 10^{-8}$.
The most sensitive limits are comparable to the
measured abundance of H$_2$O in the Galactic star forming region W3
(Wilson et al. \cite{w03}).
However, if only 10\% of the molecular gas is in very dense
cores, then the water abundance limits in the cores would be larger
by a factor of 10 i.e. $2\times 10^{-8}$ to $1\times 10^{-7}$.
Similarly, if the average density in these starburst galaxies is
$10^4-10^5$ cm$^{-3}$, the abundance limits would be larger by factors
of 10-100.

These observations can provide useful guidance for planning 
additional searches
for H$_2$O emission with future space missions.
The most sensitive 3$\sigma$ upper limit to the H$_2$O emission obtained with
Odin is 0.5 K km s$^{-1}$ (for IC342 and M51), 
which corresponds to 1600 Jy km s$^{-1}$
or $3\times 10^{-17}$ W m$^{-2}$. The currently quoted sensitivity for the
HIFI 460-680 GHz heterodyne spectrometer\footnote{Calculated using
  HSPOT V2.0.0 and assuming an aperture efficiency of 70\%.} on the
Herschel Space Observatory 
corresponds to a 3$\sigma$ upper limit of $1.4 \times 10^{-18}$
W m$^{-2}$ for a spectral line of width 130 km s$^{-1}$ for one hour
of dual beam switched observations, or about
twenty times more sensitive than our existing Odin data.
Thus, to obtain H$_2$O spectra of these 
galaxies which are
ten times more sensitive than the Odin spectra presented here will require
integration times with Herschel of only 5-20 minutes per source.

Alternatively, we can estimate what integration time will be needed to reach an
H$_2$O abundance limit of $2\times 10^{-10}$, e.g., a factor of 10
lower than the abundance of water in W3 (Wilson et al. \cite{w03}).
For NGC253 and M82, this limit is reached in 3-6 minutes, while
IC342 and M51 would require $\sim 30$ 
minutes of integration. However, CenA and NGC4258
require 3.5 and  5.1 hours, respectively, to reach this abundance limit.
These longer integration times are primarily due to the lower 
surface density of molecular gas in these two galaxies.
This analysis suggests that significant integration times may be required
to detect H$_2$O in emission even with Herschel, and that the best targets
for the initial searches will be the most gas-rich, nearby starburst
galaxies. 
If deep searches are not successful in detecting H$_2$O in emission,
it may have interesting implications for
the astrochemistry of starburst galaxies. However,
the recent ISO results on H$_2$O absorption lines in starburst galaxies
(Goicoechea et al. \cite{g05}, Gonzalez-Alfonso et al. \cite{g04})
suggest that absorption line searches with Herschel may be very fruitful,
especially given its much improved
angular resolution compared to Odin.

\begin{acknowledgements}

Generous financial support from the Research Councils
and Space Agencies in Canada, Finland, France, and Sweden
is gratefully acknowledged.  We thank the referee for useful comments.
\end{acknowledgements}


\end{document}